\documentclass[showpacs,aps,prl,twocolumn,superscriptaddress,longbibliography]{revtex4-1}
\usepackage{graphicx}
\usepackage{bm}
\usepackage{color}
\usepackage{amsmath}
\usepackage{amssymb}
\usepackage{enumerate}
\usepackage{xspace}
\usepackage{mathrsfs}
\usepackage{mathptmx}
\usepackage{graphicx}
\usepackage{amsmath,amssymb}

\newcommand{\Neel}{N\'eel }
\usepackage[normalem]{ulem}

\begin{document}

\newcommand{\Ham}{\ensuremath{\mathscr{H}}\xspace}
\newcommand{\ku}{\ensuremath{k_{\mathrm{u}}}\xspace}
\newcommand{\Sz}{\ensuremath{S_z}}
\newcommand{\Jij}{\ensuremath{J_{ij}}\xspace}
\newcommand{\mus}{\ensuremath{\mu_{\mathrm{s}}}\xspace}
\newcommand{\sH}{\ensuremath{\mathbf{H}}\xspace}
\newcommand{\Happ}{\ensuremath{\mathbf{H}_{\mathrm{app}}}\xspace}
\newcommand{\sS}{\ensuremath{\mathbf{S}}\xspace}
\newcommand{\TC}{\ensuremath{T_{\mathrm{C}}}\xspace}
\newcommand{\TN}{\ensuremath{T_{\mathrm{N}}}\xspace}
\newcommand{\kB}{\ensuremath{k_{\mathrm{B}}}\xspace}
\newcommand{\muB}{\ensuremath{\mu_{\mathrm{B}}}\xspace}
\newcommand{\factor}{3.92\xspace}
\newcommand{\vampire}{\textsc{vampire}\xspace}
\newcommand{\abinitio}{\textit{ab-initio}\xspace}
\newcommand{\AFM}{\textsc{afm}\xspace}
\newcommand{\FCC}{\textsc{fcc}\xspace}
\newcommand{\etal}{\textit{et al}\xspace}

\newcommand{\kN}{\ensuremath{k_{ij}}\xspace}
\newcommand{\eij}{\ensuremath{\mathbf{e}_{ij}}\xspace}
\newcommand{\Jnn}{\ensuremath{J_{ij}^{\mathrm{nn}}}\xspace}
\newcommand{\Jnnn}{\ensuremath{J_{ij}^{\mathrm{nnn}}}\xspace}

\newcommand{\IrMn}{\ensuremath{\text{IrMn}_3}\xspace}
\newcommand{\MnAu}{\ensuremath{\text{Mn}_2\text{Au}}\xspace}
\newcommand{\Lonetwo}{\ensuremath{\text{L}1_2}\xspace}

\title{Extraordinary temperature dependent magnetic anisotropy \\ of the non-collinear antiferromagnet \IrMn }
\author{Sarah Jenkins}
\email{sarah.jenkins@york.ac.uk}
\affiliation{Department of Physics, The University of York, York, YO10 5DD, UK}
\author{Roy Chantrell}
\email{roy.chantrell@york.ac.uk}
\affiliation{Department of Physics, The University of York, York, YO10 5DD, UK}
\author{Timothy Klemmer}
\affiliation{Seagate Research, Fremont, California 94538, USA}
\author{Richard F. L. Evans}
\email{richard.evans@york.ac.uk}
\affiliation{Department of Physics, The University of York, York, YO10 5DD, UK}

\begin{abstract}
The magnetic anisotropy of antiferromagnets plays a crucial role in stabilising the magnetisation of many spintronic devices. In non-collinear antiferromagnets such as IrMn the symmetry and temperature dependence of the effective anisotropy are poorly understood. Theoretical and experimental calculations of the effective anisotropy constant for IrMn differ by two orders of magnitude, while the symmetry has been calculated as uniaxial in contradiction to the assumed relationship between crystallographic symmetry and temperature dependence of the anisotropy from the Callen-Callen law. In this letter we determine the effective anisotropy energy surface of \Lonetwo - \IrMn using an atomistic spin model and constrained Monte Carlo simulations. We find that meta-stable spin structures lower the overall energy barrier to a tenth of that estimated from simple geometrical considerations, significantly reducing the discrepancy between experiment and theory. The temperature scaling of the anisotropy energy barrier shows an exponent of \factor, close to a uniaxial exponent of 3. Our results demonstrate the importance of non-collinear spin states on the thermal stability of antiferromagnets with consequences for the practical application of antiferromagnets in devices operating at elevated temperatures.
\end{abstract}

\maketitle


\textit{Introduction -} The magnetic anisotropy of antiferromagnetic materials plays a key role in the stability of many spintronic devices~\cite{Wadley2016ElectricalAntiferromagnet,Jungwirth2018TheSpintronics,Jungwirth2016AntiferromagneticSpintronics,Meinert2017ElectricalActivation,Lin2019} and exchange bias effects in general~\cite{Meiklejohn1957NewAnisotropy, OGrady2010AFilms, Nogues1999ExchangeBias}. Recently, interest in the properties of antiferromagnetic materials has increased due to their emerging applications in antiferromagnetic spintronic~\cite{Jungwirth2016AntiferromagneticSpintronics,Lin2019} and neuromorphic computing devices~\cite{Ielmini2018Brain-inspiredNetworks} where the antiferromagnet is the active element. The magnetic anisotropy of antiferromagnets is poorly understood due to the difficulty in experimental measurements and the complexity of the materials. Iridium Manganese (IrMn) is the material of choice for many spintronic devices due to its high thermal stability and large exchange bias. In devices the ordering and composition is tuned for optimal performance but here we focus on the \Lonetwo ordered - \IrMn phase due to the existence of extensive experimental~\cite{Kohn2013TheExchange-bias.,Tomeno1999MagneticMn3Ir} and theoretical~\cite{Szunyogh2009GiantPrinciples, Szunyogh2011AtomisticInterface,Tsunoda2007UncompensatedBilayers} data.

Theoretical simulations by Szunyogh~\textit{et al} \cite{Szunyogh2009GiantPrinciples} found an extremely large second order magnetic anisotropy for \IrMn, leading to a predicted magnetocrystalline  anisotropy energy density (MAE) of the order of $3 \times 10^7$ J/m$^3$ at 0 K. Vallejo-Fernandez \textit{et al}~\cite{Vallejo-Fernandez2007MeasurementSystems} experimentally determined the anisotropy constant of IrMn by measuring the mean blocking temperature of an IrMn/CoFe bilayer. Their measurements used a training-free measurement procedure in which hysteresis loops were repeatedly measured at the same (thermal activation free) low temperature after raising the sample to an increased activation temperature. The activation reverses part of the AF layer due to the exchange field from the ferromagnet. During this procedure, the exchange bias field changes sign at the blocking temperature, from which the anisotropy can be determined given the measured grain volume ~\cite{Vallejo-Fernandez2007MeasurementSystems,Carpenter2014EffectNanostructures}. Using this procedure, Carpenter \textit{et al}~\cite{Carpenter2014EffectNanostructures} calculated a value of the MAE of 6.5 $\times$ 10$^5$ J/m$^3$  at 300K. This value is almost two orders of magnitude lower than the theoretical calculation~\cite{Szunyogh2009GiantPrinciples}. The experimental estimate of the anisotropy constant of IrMn is sensitive to the value of the switching attempt frequency ($f_0$) in the Arrhenius \Neel law given by: 
\begin{equation}
1/\tau = f_0 \exp \left(-\frac{\Delta{E}}{k_BT}\right)
\label{eq:switch}
\end{equation}
where $\tau$ is the relaxation time, $\Delta{E}$ is the energy barrier, $k_B$ is the Boltzmann constant, and $T$ is the temperature. Originally Vallejo-Fernandez \etal used a value of $f_0 = 10^9$~s$^{-1}$ \cite{Vallejo-Fernandez2007MeasurementSystems} but more recent estimates suggest values closer to $f_0 = 10^{12}$~s$^{-1}$~\cite{Vallejo-Fernandez2010MeasurementAntiferromagnets}.

A further unresolved problem relates to the symmetry of the \IrMn anisotropy. The measurement procedure of Ref.~\cite{Vallejo-Fernandez2007MeasurementSystems} requires measurements over a large range of temperatures, which are strongly affected by the temperature variation of the anisotropy. Fitting to the experiments by Vallejo-Fernandez ~\cite{Vallejo-Fernandez2007MeasurementSystems} and Craig \textit{et al}~\cite{Craig2008ABilayers} used a Callen-Callen-type \cite{Callen1966TheLaw} power law $K_{AF}(T)/K_{AF}(0) = (n_{AF}(T)/n_{AF}(0))^l$, with $n_{AF}$ the AF sublattice magnetisation. The exponent $l$ reflects the symmetry of the anisotropy, which itself generally reflects that of the lattice. Agreement with experimental measurements~\cite{Vallejo-Fernandez2007MeasurementSystems} requires an exponent of $l \sim 3$ corresponding to uniaxial rather than cubic anisotropy, which would give $l=10$. Szunyogh \etal~\cite{Szunyogh2009GiantPrinciples} showed that the local energy surface for individual spins is uniaxial by rotating the triangular ground state about the (111) direction. Both experiment and theory agree that the anisotropy has an approximately  uniaxial form as inferred from the temperature scaling in experiments and from the site symmetry in the theory. However, this contradicts the predicted relationship between crystallographic symmetry and the temperature dependence of the anisotropy from the Callen-Callen law~\cite{Callen1966TheLaw}. 

\textit{Method -} To study the anisotropy of \Lonetwo - \IrMn we use an atomistic spin model where the energy of the system is defined using the spin Hamiltonian:
\begin{equation}
\Ham = -\sum_{i<j} \Jij \sS_i \cdot \sS_j - \frac{k_N}{2} \sum_{i \neq j}^z (\mathbf{S}_i \cdot  \mathbf{e}_{ij})^2 
\label{eq:hamiltonian}
\end{equation}
where $\sS_i$ is a unit vector describing spin direction on Mn site i, $k_N = -4.22 \times 10^{-22}$ is the \Neel anisotropy constant and $\mathbf{e}_{ij}$ is a unit vector from site $i$ to site $j$, $z$ is the number of nearest neighbours and \Jij is the exchange interaction. The effective exchange interactions (\Jij) were limited to nearest ($\Jnn = -6.4 \times 10^{-21}$ J/link) and next nearest ($\Jnnn = 5.1 \times 10^{-21}$ J/link) neighbours~\cite{Jenkins2018EnhancedFilms}. In IrMn the magnetocrystalline anisotropy arises from the large spin-orbit coupling between Mn and Ir sites~\cite{Szunyogh2009GiantPrinciples}. Here we map the local anisotropies at each Mn site to a \Neel pair anisotropy model~\cite{Neel1954LapprocheMagnetostriction, Jenkins2018EnhancedFilms} which gives exact agreement with the \emph{ab initio} calculations given in ref~\cite{Szunyogh2009GiantPrinciples}. The \Neel model reflects the local site symmetry to give the correct easy axes for each Mn site and by performing coherent spin rotations as in \cite{Szunyogh2009GiantPrinciples} we find the same angular dependence of the anisotropy energy.

\begin{figure}[tb]
\centering
\includegraphics[width=70mm]{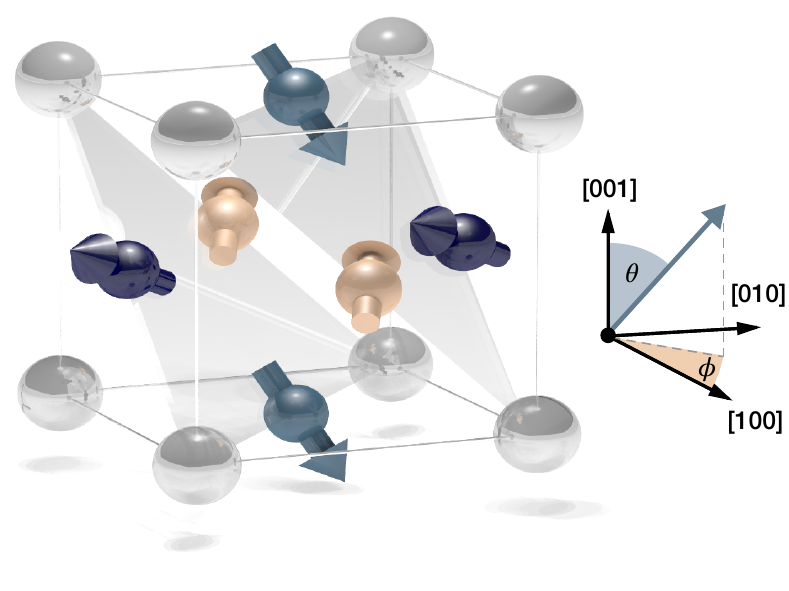}
\caption{Visualisation of the simulated ground state spin structure of \Lonetwo-\IrMn obtained from zero-field cooling. The spin directions show an average spin of each magnetic sublattice direction over the whole sample. The corner atoms represent Ir and so have no net magnetic moment. The simulated spin structure agrees with experimental measurements and first principles simulations. Crystallographic directions and reference directions for constraint angles ($\theta$, $\phi$) for the sublattice magnetisation are shown inset.}
\label{fig:system}
\end{figure}

\textit{Results -} To verify the model we calculated the ground state spin structure of ordered $\Lonetwo$ \IrMn using a Monte Carlo Metropolis algorithm with the adaptive update method~\cite{Alzate-Cardona2019OptimalSystems,Evans2014} and implemented in the \textsc{vampire} software package~\cite{vmpr}. The $8 \times 8 \times 8$ nm$^3$ system was initially equilibrated at a temperature of 1500 K (above the \Neel temperature) to thermalise the spins. The system was then cooled to 0 K using a linear cooling function over 10$^6$ Monte Carlo steps to find a ground state spin configuration. In agreement with previous experimental~\cite{Tomeno1999MagneticMn3Ir,Kohn2013TheExchange-bias.} and \abinitio results~\cite{Szunyogh2009GiantPrinciples} we find that ordered \Lonetwo-\IrMn  has a triangular ($T1$) spin structure where the magnetic moments lie parallel to the [111] planes as shown in Fig.~\ref{fig:system}. There are 8 possible [111] planes and by symmetry IrMn therefore has 8 magnetic ground states.  

\begin{figure}[tb]
\centering
\includegraphics[width=80mm]{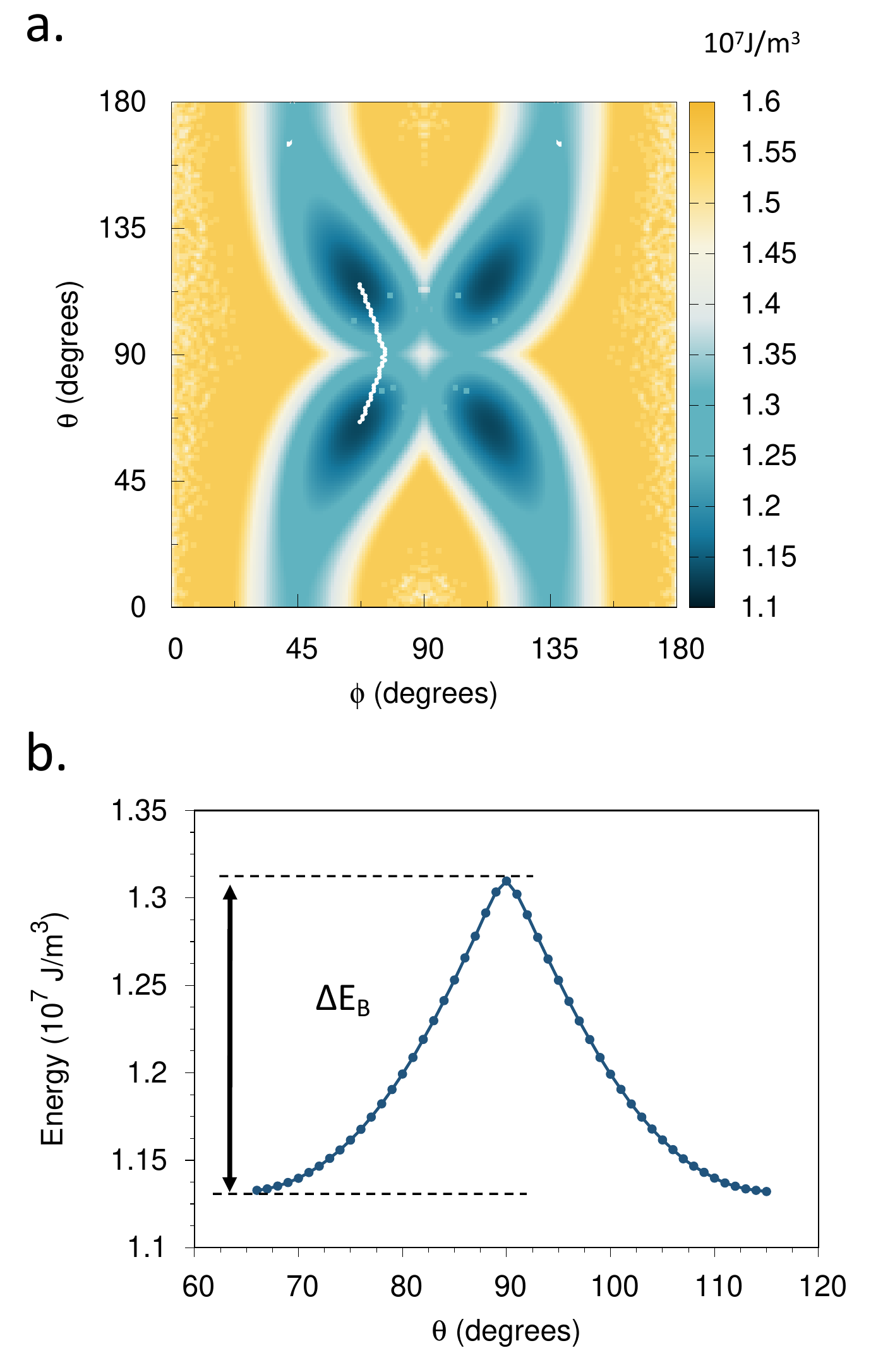}
\caption{(a) Simulated anisotropy energy surface for ordered \Lonetwo - \IrMn recovered from the integral of the total torque given by Eq.~4. (b) Cross section of the anisotropy surface showing the minimum energy path to reversal. The energy barrier $\Delta E_{\mathrm{B}}$ to move between the minima is shown. Colour Online.}
\label{fig:energy}
\end{figure}

The energy barrier separating two ground states is the minimum energy path for the spins to rotate between them. The energy barrier defines the effective MAE and therefore the thermal stability of \IrMn. To calculate this we use the constrained Monte Carlo algorithm to determine the energy surface and the energy barrier to magnetic reversal\cite{Asselin2010ConstrainedAnisotropy}. Here, we constrain the direction of magnetisation of a single Mn sublattice while allowing all other spins to relax to obtain the ground state structure with a constraint applied. By scanning all angles ($\theta,\phi$) the energy surface is obtained. For each value of  $\theta$ and $\phi$ the  (8 nm)$^3$ system was initially heated to 1500K  to thermalize the spins and then cooled to 0K. Due to the constraint the system cannot reach a full equilibrium and so the total internal torque ($\mathbf{\tau}$) is non-zero and given by
\begin{equation}
\mathbf{\tau} = -\mathbf{M} \times \frac{\partial \mathscr{F}} {\partial\mathbf{M}}
\end{equation}
where $\mathscr{F}$ is the Helmholtz free energy which is a function of $\textbf{M}$. Since $\mathscr{F}$ cannot be computed directly we reconstruct it from the integral of the torque
\begin{equation}
\mathscr{F} = \mathscr{F}_0 + \int^{M'}_M (\textbf{M}' \times \textbf{T}) \cdot d\textbf{M}'
\end{equation} 
by numerical integration of the torque taken along any path between two points on the energy surface. The computed energy surface at 0K is shown in Fig. \ref{fig:energy}(a) and has a complicated structure with four minima. The energy minima lie at $\phi \sim \pm 24^{\circ}$ corresponding to the expected easy directions. 
To calculate the energy barrier between two adjacent minima we compute the minimum energy path between them as shown in Fig.~\ref{fig:energy}(b). The calculated 0K energy barrier is $1.17 \times 10^6$ J/m$^3$ which is an order of magnitude lower than that for rigid rotation of spins calculated by Szunyogh \etal~\cite{Szunyogh2009GiantPrinciples} but still an order of magnitude more than the experimental measurement. The surprising reduction arises due to a bobbing effect where the exchange and anisotropy energies compete to lower the total energy through small deviations from the ground-state spin structure when the antiferromagnetic spins are rotated. This is particularly relevant to macroscopic approximations of antiferromagnetic materials with \Neel vectors where the sublattices are always assumed to have a fixed local spin structure. In the case of IrMn the exceptionally high magnetic anisotropy means that this approximation is not applicable away from the ground state and would lead to an overestimation of the energy barrier to switching. We note that, although the energy surface illustrated in Fig.~\ref{fig:energy}(a) has an unusually complex form, the minima themselves exhibit a four-fold symmetry, characteristic of cubic rather than uniaxial anisotropy. The question is: how to resolve the apparent contradiction with the experimental data of Vallejo-Fernandez \etal~\cite{Vallejo-Fernandez2007MeasurementSystems} and its requirement of a magnetisation scaling exponent consistent with uniaxial symmetry.

\begin{figure}[tb]
\centering
\includegraphics[width=80mm]{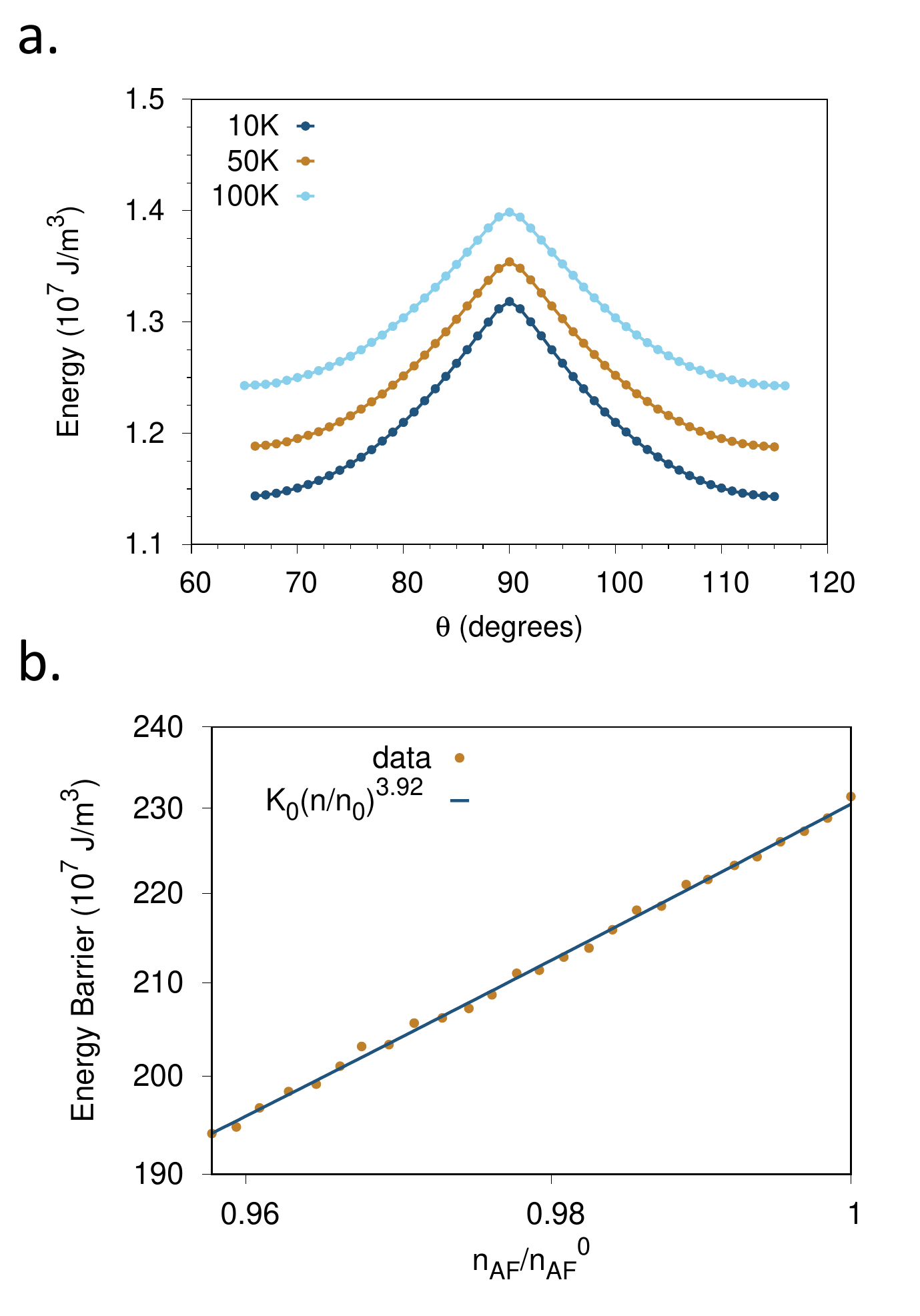}
\caption{Simulated temperature dependence of the energy barrier shown by minimum energy paths (a) The energy surface for temperatures of 0K, 100K and 300K. The total anisotropy energy increases due to spin fluctuations, but the energy barrier decreases with temperature. (b) The scaling of the effective energy barrier with sublattice magnetisation length $n_{\mathrm{AF}}$ fitted using $E_B(n_{\mathrm{AF}}) = E_0 n_{\mathrm{AF}}^{l}$. $l$ is calculated to be $\factor \pm 12$  suggesting a scaling much slower than cubic anisotropy where $l = 10$ and more similar to uniaxial anisotropy $l = 3$.}
\label{fig:temperature}
\end{figure}

To resolve this discrepancy we now investigate the temperature dependence of the energy barrier theoretically to calculate the scaling exponent. The energy surfaces and minimum energy path were calculated for temperatures between 0K and 350K as shown in Fig.~\ref{fig:temperature}(a). The absolute anisotropy energy increases with temperature due to spin fluctuations but the energy barrier between neighbouring ground states decreases, characteristic of a reduction of the anisotropy. In Fig.~\ref{fig:temperature}(b) we plot the power law dependence of the effective energy barrier as a function of the magnetisation and find a unusual exponent of $l = \factor \pm 0.14$. The exponent is close to a uniaxial exponent of $l = 3$ matching the experimental observations but deviates from this ideal value due to the complex symmetry of the anisotropy energy surface. We conclude that the magnetic anisotropy of \Lonetwo - \IrMn possesses a close to uniaxial temperature dependence in direct contradiction with the usual Callen - Callen power laws and cubic nature of the crystal \cite{Callen1966TheLaw}.

The attempt frequency $f_0$ is a critical parameter in calculating the effective anisotropy of antiferromagnets from experimental data. Having determined the precise energy barrier at an elevated temperature we are now able to compute the attempt frequency using atomistic spin dynamics. We simulate the dynamic behaviour using the stochastic Landau–Lifshitz-Gilbert (sLLG) equation~\cite{Evans2014,Ellis} given by
\begin{equation}
\frac{\partial \mathbf{S}_i}{\partial t} = -\frac{\gamma }{1 + \lambda^2} \left[\mathbf{S}_i \times \mathbf{B}_{\mathrm{eff}} + \lambda \mathbf{S}_i\times (\mathbf{S}_i \times  \mathbf{B}_{\mathrm{eff}} )\right],
\end{equation}
where the Gilbert damping constant $\lambda = 0.1$ and $\gamma$ is the absolute value of the gyromagnetic ratio. The effective field $\mathbf{B}_{\mathrm{eff}}$  is calculated as the derivative of the spin Hamiltonian with respect to the local spin moment plus a random thermal field ($\mathbf{B}_{\mathrm{eff}} = -{\mu_S}^{-1} \partial \mathbf{\mathscr{H}}/ \partial \mathbf{S}_i+ \mathbf{B}_{\mathrm{th}}^i$) where $\mathbf{B}_{\mathrm{th}}^i = \Gamma(t) \sqrt{\frac{2\lambda k_B T}{\gamma \mu_S \Delta t}}$ and $\Gamma$ is a 3D random Gaussian distribution. The sLLG equation is integrated using a second order predictor corrector Heun scheme \cite{Evans2014}. 

\begin{figure}[tbh]
\centering
\includegraphics[width=80mm]{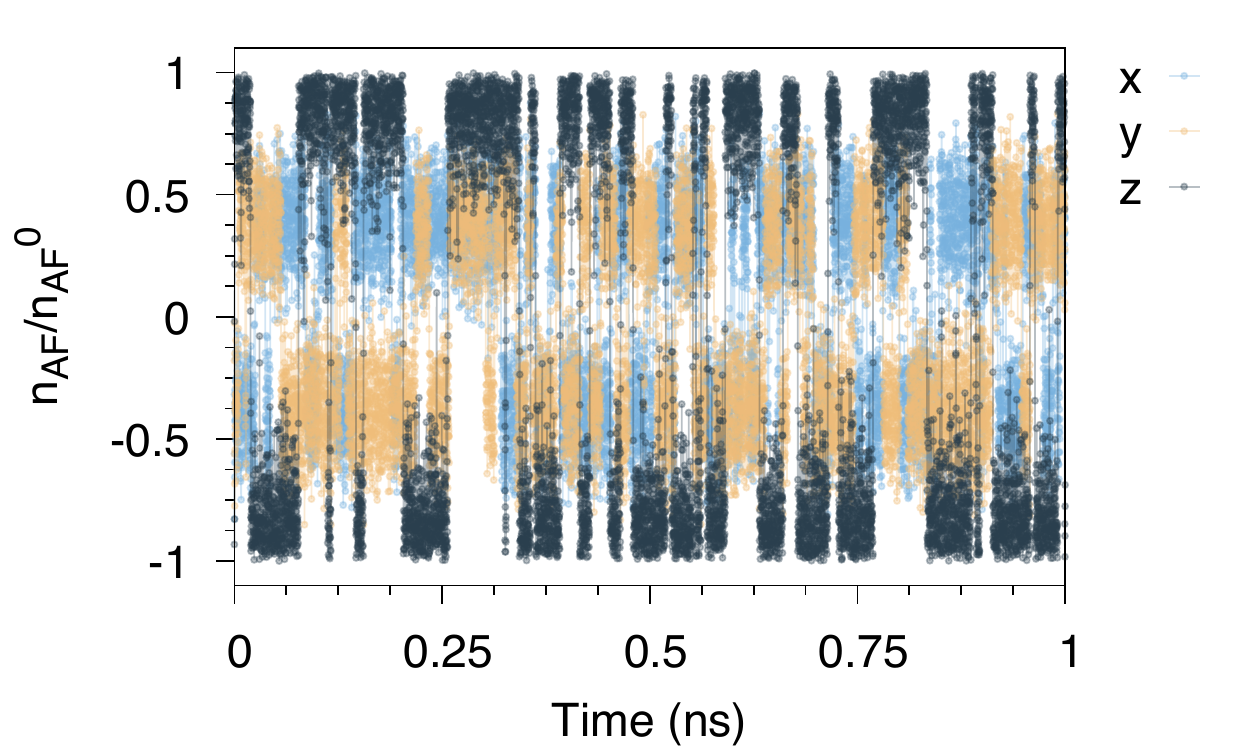}
\caption{Time-dependent magnetisation of \IrMn at 100K simulated using atomistic spin dynamics. The simulation was run for 100 ns, the first 1 ns is shown here. The sublattice magnetisation flips superparamagnetically between different coherent ground state orientations. At this temperature the sublattice ordering is approximately 90\% since the system is simulated far from the \Neel temperature.}
\label{fig:switch}
\end{figure}

We determined the attempt frequency by calculating the transition rate just below the blocking temperature of the antiferromagnet. Due to the giant anisotropy of IrMn and limited time accessible by simulations we simulate a small sample only (1.5 nm)$^3$ which has a blocking temperature of $T_B = 101.5$K for a timescale of 0.1 ns. To get a precise estimate of the energy barrier for this system size at 100K we use the same method as above to calculate the energy surface. The time dependent dynamics of the magnetisation for a single sublattice is shown in Fig.~\ref{fig:switch} over a simulation time of one nanosecond.
As the temperature is just below the blocking temperature the IrMn switches between stable states giving a time dependent form similar to telegraph noise. Over a total simulation time of 100 ns the total number of switches was calculated and divided by the total simulation time. This gave an average time between switches of $4.03 \times 10^{-11}$ s at $T = 100$ K. Using Eq.~\ref{eq:switch} and the energy barrier calculated at 100K we find $f_0$ to be $1.25 \times 10^{13}$ Hz. This is somewhat higher than the experimentally calculated value \cite{Vallejo-Fernandez2010MeasurementAntiferromagnets}. Using this value for the attempt frequency increases the experimental value of the anisotropy energy barrier by a factor of two to $15.73 \times 10^5$ J/m$^3$ at 0K. The combination of a lower energy barrier and higher attempt frequency reduces the difference between the theoretical and experimental calculations to only a factor two. The remaining difference in the values of the effective magnetic anisotropy could be due to different ordering or defects in the experimental samples, but our results finally resolve the large disparity between the theoretically calculated and experimentally measured magnetic anisotropy of \IrMn.

\textit{Discussion -}
Applying constrained minimisation and spin dynamics simulations we have determined the effective temperature dependent anisotropy and relaxation dynamics of \IrMn, one of the most technologically important non-collinear antiferromagnetic materials. We find that the anisotropic energy surface is unusually complex and find a scaling exponent of the effective magnetic anisotropy that is fundamentally different from the usual Callen-Callen theory despite the presence of cubic crystal symmetry and localised uniaxial anisotropy at atomic Mn sites. Meta-stable spins structures are shown to lower the overall energy barrier to a tenth of that estimated from simple geometrical approximations. Spin dynamics calculations reveal an exceptionally high attempt frequency in \IrMn of $f_0 = 1.25 \times 10^{13}$ s$^{-1}$; a value four orders of magnitude larger than the typical value for ferromagnets of $10^9$ s$^{-1}$. The combination of a lower energy barrier and higher attempt frequency reduces the discrepancy between theory and experiment by over an order of magnitude. We conclude that the magnetic anisotropy of \Lonetwo- \IrMn possesses a close to uniaxial temperature dependence but different from that expected for either pure uniaxial or cubic anisotropy expected from the usual Callen-Callen relation. 

Although we have focused on \IrMn we expect that other non-collinear antiferromagnets such as MnPt and MnFe will exhibit similarly complex temperature dependent magnetic anisotropy. This is likely to be strongly affected by composition and ordering which will disrupt the local anisotropy energy surface at different atomic sites. Our results have important consequences for applications of antiferromagnets in determining their thermal stability and dynamic properties and provide an established methodology for determining the effective magnetic anisotropy at elevated temperatures. This is particularly important for emerging applications in neuromorphic computing and antiferromagnetic spintronics where the long-term stability of the antiferromagnet is critical to device operation. Further investigation may yield different classes of antiferromagnets with unusual temperature dependent properties.

\section{acknowledgements}
We gratefully acknowledge the provision of computer time made available on the \textsc{viking} cluster, a high performance compute facility provided by the University of York.

\bibliography{references,local} 

\begin{thebibliography}{26}%
\makeatletter
\providecommand \@ifxundefined [1]{%
 \@ifx{#1\undefined}
}%
\providecommand \@ifnum [1]{%
 \ifnum #1\expandafter \@firstoftwo
 \else \expandafter \@secondoftwo
 \fi
}%
\providecommand \@ifx [1]{%
 \ifx #1\expandafter \@firstoftwo
 \else \expandafter \@secondoftwo
 \fi
}%
\providecommand \natexlab [1]{#1}%
\providecommand \enquote  [1]{``#1''}%
\providecommand \bibnamefont  [1]{#1}%
\providecommand \bibfnamefont [1]{#1}%
\providecommand \citenamefont [1]{#1}%
\providecommand \href@noop [0]{\@secondoftwo}%
\providecommand \href [0]{\begingroup \@sanitize@url \@href}%
\providecommand \@href[1]{\@@startlink{#1}\@@href}%
\providecommand \@@href[1]{\endgroup#1\@@endlink}%
\providecommand \@sanitize@url [0]{\catcode `\\12\catcode `\$12\catcode
  `\&12\catcode `\#12\catcode `\^12\catcode `\_12\catcode `\%12\relax}%
\providecommand \@@startlink[1]{}%
\providecommand \@@endlink[0]{}%
\providecommand \url  [0]{\begingroup\@sanitize@url \@url }%
\providecommand \@url [1]{\endgroup\@href {#1}{\urlprefix }}%
\providecommand \urlprefix  [0]{URL }%
\providecommand \Eprint [0]{\href }%
\providecommand \doibase [0]{http://dx.doi.org/}%
\providecommand \selectlanguage [0]{\@gobble}%
\providecommand \bibinfo  [0]{\@secondoftwo}%
\providecommand \bibfield  [0]{\@secondoftwo}%
\providecommand \translation [1]{[#1]}%
\providecommand \BibitemOpen [0]{}%
\providecommand \bibitemStop [0]{}%
\providecommand \bibitemNoStop [0]{.\EOS\space}%
\providecommand \EOS [0]{\spacefactor3000\relax}%
\providecommand \BibitemShut  [1]{\csname bibitem#1\endcsname}%
\let\auto@bib@innerbib\@empty
\bibitem [{\citenamefont {Wadley}\ \emph {et~al.}(2016)\citenamefont {Wadley},
  \citenamefont {Howells}, \citenamefont { elezny}, \citenamefont {Andrews},
  \citenamefont {Hills}, \citenamefont {Campion}, \citenamefont {Novak},
  \citenamefont {Olejnik}, \citenamefont {Maccherozzi}, \citenamefont {Dhesi},
  \citenamefont {Martin}, \citenamefont {Wagner}, \citenamefont {Wunderlich},
  \citenamefont {Freimuth}, \citenamefont {Mokrousov}, \citenamefont {Kune },
  \citenamefont {Chauhan}, \citenamefont {Grzybowski}, \citenamefont
  {Rushforth}, \citenamefont {Edmonds}, \citenamefont {Gallagher},\ and\
  \citenamefont {Jungwirth}}]{Wadley2016ElectricalAntiferromagnet}%
  \BibitemOpen
  \bibfield  {author} {\bibinfo {author} {\bibfnamefont {P.}~\bibnamefont
  {Wadley}}, \bibinfo {author} {\bibfnamefont {B.}~\bibnamefont {Howells}},
  \bibinfo {author} {\bibfnamefont {J.}~\bibnamefont { elezny}}, \bibinfo
  {author} {\bibfnamefont {C.}~\bibnamefont {Andrews}}, \bibinfo {author}
  {\bibfnamefont {V.}~\bibnamefont {Hills}}, \bibinfo {author} {\bibfnamefont
  {R.~P.}\ \bibnamefont {Campion}}, \bibinfo {author} {\bibfnamefont
  {V.}~\bibnamefont {Novak}}, \bibinfo {author} {\bibfnamefont
  {K.}~\bibnamefont {Olejnik}}, \bibinfo {author} {\bibfnamefont
  {F.}~\bibnamefont {Maccherozzi}}, \bibinfo {author} {\bibfnamefont {S.~S.}\
  \bibnamefont {Dhesi}}, \bibinfo {author} {\bibfnamefont {S.~Y.}\ \bibnamefont
  {Martin}}, \bibinfo {author} {\bibfnamefont {T.}~\bibnamefont {Wagner}},
  \bibinfo {author} {\bibfnamefont {J.}~\bibnamefont {Wunderlich}}, \bibinfo
  {author} {\bibfnamefont {F.}~\bibnamefont {Freimuth}}, \bibinfo {author}
  {\bibfnamefont {Y.}~\bibnamefont {Mokrousov}}, \bibinfo {author}
  {\bibfnamefont {J.}~\bibnamefont {Kune }}, \bibinfo {author} {\bibfnamefont
  {J.~S.}\ \bibnamefont {Chauhan}}, \bibinfo {author} {\bibfnamefont {M.~J.}\
  \bibnamefont {Grzybowski}}, \bibinfo {author} {\bibfnamefont {A.~W.}\
  \bibnamefont {Rushforth}}, \bibinfo {author} {\bibfnamefont {K.~W.}\
  \bibnamefont {Edmonds}}, \bibinfo {author} {\bibfnamefont {B.~L.}\
  \bibnamefont {Gallagher}}, \ and\ \bibinfo {author} {\bibfnamefont
  {T.}~\bibnamefont {Jungwirth}},\ }\bibfield  {title} {\enquote {\bibinfo
  {title} {{Electrical switching of an antiferromagnet}},}\ }\href {\doibase
  10.1126/science.aab1031} {\bibfield  {journal} {\bibinfo  {journal}
  {Science}\ }\textbf {\bibinfo {volume} {351}},\ \bibinfo {pages} {587--590}
  (\bibinfo {year} {2016})}\BibitemShut {NoStop}%
\bibitem [{\citenamefont {Jungwirth}\ \emph {et~al.}(2018)\citenamefont
  {Jungwirth}, \citenamefont {Sinova}, \citenamefont {Manchon}, \citenamefont
  {Marti}, \citenamefont {Wunderlich},\ and\ \citenamefont
  {Felser}}]{Jungwirth2018TheSpintronics}%
  \BibitemOpen
  \bibfield  {author} {\bibinfo {author} {\bibfnamefont {T.}~\bibnamefont
  {Jungwirth}}, \bibinfo {author} {\bibfnamefont {J.}~\bibnamefont {Sinova}},
  \bibinfo {author} {\bibfnamefont {A.}~\bibnamefont {Manchon}}, \bibinfo
  {author} {\bibfnamefont {X.}~\bibnamefont {Marti}}, \bibinfo {author}
  {\bibfnamefont {J.}~\bibnamefont {Wunderlich}}, \ and\ \bibinfo {author}
  {\bibfnamefont {C.}~\bibnamefont {Felser}},\ }\bibfield  {title} {\enquote
  {\bibinfo {title} {{The multiple directions of antiferromagnetic
  spintronics}},}\ }\href {\doibase 10.1038/s41567-018-0063-6} {\bibfield
  {journal} {\bibinfo  {journal} {Nature Physics}\ }\textbf {\bibinfo {volume}
  {14}},\ \bibinfo {pages} {200--203} (\bibinfo {year} {2018})}\BibitemShut
  {NoStop}%
\bibitem [{\citenamefont {Jungwirth}\ \emph {et~al.}(2016)\citenamefont
  {Jungwirth}, \citenamefont {Marti}, \citenamefont {Wadley},\ and\
  \citenamefont {Wunderlich}}]{Jungwirth2016AntiferromagneticSpintronics}%
  \BibitemOpen
  \bibfield  {author} {\bibinfo {author} {\bibfnamefont {T.}~\bibnamefont
  {Jungwirth}}, \bibinfo {author} {\bibfnamefont {X.}~\bibnamefont {Marti}},
  \bibinfo {author} {\bibfnamefont {P.}~\bibnamefont {Wadley}}, \ and\ \bibinfo
  {author} {\bibfnamefont {J.}~\bibnamefont {Wunderlich}},\ }\bibfield  {title}
  {\enquote {\bibinfo {title} {{Antiferromagnetic spintronics}},}\ }\href
  {\doibase 10.1038/nnano.2016.18} {\bibfield  {journal} {\bibinfo  {journal}
  {Nature Nanotechnology}\ }\textbf {\bibinfo {volume} {11}},\ \bibinfo {pages}
  {231--241} (\bibinfo {year} {2016})}\BibitemShut {NoStop}%
\bibitem [{\citenamefont {Meinert}\ \emph {et~al.}(2017)\citenamefont
  {Meinert}, \citenamefont {Graulich},\ and\ \citenamefont
  {Matalla-Wagner}}]{Meinert2017ElectricalActivation}%
  \BibitemOpen
  \bibfield  {author} {\bibinfo {author} {\bibfnamefont {Markus}\ \bibnamefont
  {Meinert}}, \bibinfo {author} {\bibfnamefont {Dominik}\ \bibnamefont
  {Graulich}}, \ and\ \bibinfo {author} {\bibfnamefont {Tristan}\ \bibnamefont
  {Matalla-Wagner}},\ }\bibfield  {title} {\enquote {\bibinfo {title}
  {{Electrical switching of antiferromagnetic Mn{\$}{\_}2{\$}Au and the role of
  thermal activation}},}\ }\href {\doibase 10.1103/PhysRevApplied.9.064040} {\
  (\bibinfo {year} {2017}),\ 10.1103/PhysRevApplied.9.064040}\BibitemShut
  {NoStop}%
\bibitem [{\citenamefont {Lin}\ \emph {et~al.}(2019)\citenamefont {Lin},
  \citenamefont {Yang}, \citenamefont {Tsai}, \citenamefont {Chen},
  \citenamefont {Huang}, \citenamefont {Lin},\ and\ \citenamefont
  {Lai}}]{Lin2019}%
  \BibitemOpen
  \bibfield  {author} {\bibinfo {author} {\bibfnamefont {Po-Hung}\ \bibnamefont
  {Lin}}, \bibinfo {author} {\bibfnamefont {Bo-Yuan}\ \bibnamefont {Yang}},
  \bibinfo {author} {\bibfnamefont {Ming-Han}\ \bibnamefont {Tsai}}, \bibinfo
  {author} {\bibfnamefont {Po-Chuan}\ \bibnamefont {Chen}}, \bibinfo {author}
  {\bibfnamefont {Kuo-Feng}\ \bibnamefont {Huang}}, \bibinfo {author}
  {\bibfnamefont {Hsiu-Hau}\ \bibnamefont {Lin}}, \ and\ \bibinfo {author}
  {\bibfnamefont {Chih-Huang}\ \bibnamefont {Lai}},\ }\bibfield  {title}
  {\enquote {\bibinfo {title} {Manipulating exchange bias by spin-orbit
  torque},}\ }\href {\doibase 10.1038/s41563-019-0289-4} {\bibfield  {journal}
  {\bibinfo  {journal} {Nature Materials}\ }\textbf {\bibinfo {volume} {18}},\
  \bibinfo {pages} {335--341} (\bibinfo {year} {2019})}\BibitemShut {NoStop}%
\bibitem [{\citenamefont {Meiklejohn}\ and\ \citenamefont
  {Bean}(1957)}]{Meiklejohn1957NewAnisotropy}%
  \BibitemOpen
  \bibfield  {author} {\bibinfo {author} {\bibfnamefont {W.~H.}\ \bibnamefont
  {Meiklejohn}}\ and\ \bibinfo {author} {\bibfnamefont {C.~P.}\ \bibnamefont
  {Bean}},\ }\bibfield  {title} {\enquote {\bibinfo {title} {{New Magnetic
  Anisotropy}},}\ }\href {\doibase 10.1103/PhysRev.105.904} {\bibfield
  {journal} {\bibinfo  {journal} {Physical Review}\ }\textbf {\bibinfo {volume}
  {105}},\ \bibinfo {pages} {904--913} (\bibinfo {year} {1957})}\BibitemShut
  {NoStop}%
\bibitem [{\citenamefont {O'Grady}\ \emph {et~al.}(2010)\citenamefont
  {O'Grady}, \citenamefont {Fernandez-Outon},\ and\ \citenamefont
  {Vallejo-Fernandez}}]{OGrady2010AFilms}%
  \BibitemOpen
  \bibfield  {author} {\bibinfo {author} {\bibfnamefont {K.}~\bibnamefont
  {O'Grady}}, \bibinfo {author} {\bibfnamefont {L.~E.}\ \bibnamefont
  {Fernandez-Outon}}, \ and\ \bibinfo {author} {\bibfnamefont {G.}~\bibnamefont
  {Vallejo-Fernandez}},\ }\bibfield  {title} {\enquote {\bibinfo {title} {{A
  new paradigm for exchange bias in polycrystalline thin films}},}\ }\href
  {\doibase 10.1016/j.jmmm.2009.12.011} {\bibfield  {journal} {\bibinfo
  {journal} {Journal of Magnetism and Magnetic Materials}\ }\textbf {\bibinfo
  {volume} {322}},\ \bibinfo {pages} {883--899} (\bibinfo {year}
  {2010})}\BibitemShut {NoStop}%
\bibitem [{\citenamefont {Nogue´s}\ and\ \citenamefont
  {Schuller}(1999)}]{Nogues1999ExchangeBias}%
  \BibitemOpen
  \bibfield  {author} {\bibinfo {author} {\bibfnamefont {J~Nogue´s}\
  \bibnamefont {Nogue´s}}\ and\ \bibinfo {author} {\bibfnamefont {Ivan~K}\
  \bibnamefont {Schuller}},\ }\href
  {https://ac.els-cdn.com/S0304885398002662/1-s2.0-S0304885398002662-main.pdf?_tid=fac5278a-2a72-4c29-8ca0-c765878f6c70&acdnat=1541181451_a9a7d093afbb28331d56c2d349050786}
  {\emph {\bibinfo {title} {Journal of Magnetism and Magnetic Materials}}},\
  \bibinfo {type} {Tech. Rep.}\ (\bibinfo {year} {1999})\BibitemShut {NoStop}%
\bibitem [{\citenamefont {Ielmini}(2018)}]{Ielmini2018Brain-inspiredNetworks}%
  \BibitemOpen
  \bibfield  {author} {\bibinfo {author} {\bibfnamefont {Daniele}\ \bibnamefont
  {Ielmini}},\ }\bibfield  {title} {\enquote {\bibinfo {title} {{Brain-inspired
  computing with resistive switching memory (RRAM): Devices, synapses and
  neural networks}},}\ }\href {\doibase 10.1016/J.MEE.2018.01.009} {\bibfield
  {journal} {\bibinfo  {journal} {Microelectronic Engineering}\ }\textbf
  {\bibinfo {volume} {190}},\ \bibinfo {pages} {44--53} (\bibinfo {year}
  {2018})}\BibitemShut {NoStop}%
\bibitem [{\citenamefont {Kohn}\ \emph {et~al.}(2013)\citenamefont {Kohn},
  \citenamefont {Kov{\'{a}}cs}, \citenamefont {Fan}, \citenamefont {McIntyre},
  \citenamefont {Ward},\ and\ \citenamefont
  {Goff}}]{Kohn2013TheExchange-bias.}%
  \BibitemOpen
  \bibfield  {author} {\bibinfo {author} {\bibfnamefont {A}~\bibnamefont
  {Kohn}}, \bibinfo {author} {\bibfnamefont {A}~\bibnamefont {Kov{\'{a}}cs}},
  \bibinfo {author} {\bibfnamefont {R}~\bibnamefont {Fan}}, \bibinfo {author}
  {\bibfnamefont {G~J}\ \bibnamefont {McIntyre}}, \bibinfo {author}
  {\bibfnamefont {R~C~C}\ \bibnamefont {Ward}}, \ and\ \bibinfo {author}
  {\bibfnamefont {J~P}\ \bibnamefont {Goff}},\ }\bibfield  {title} {\enquote
  {\bibinfo {title} {{The antiferromagnetic structures of IrMn3 and their
  influence on exchange-bias.}}}\ }\href {\doibase 10.1038/srep02412}
  {\bibfield  {journal} {\bibinfo  {journal} {Scientific reports}\ }\textbf
  {\bibinfo {volume} {3}},\ \bibinfo {pages} {2412} (\bibinfo {year}
  {2013})}\BibitemShut {NoStop}%
\bibitem [{\citenamefont {Tomeno}\ \emph {et~al.}(1999)\citenamefont {Tomeno},
  \citenamefont {Fuke}, \citenamefont {Iwasaki}, \citenamefont {Sahashi},\ and\
  \citenamefont {Tsunoda}}]{Tomeno1999MagneticMn3Ir}%
  \BibitemOpen
  \bibfield  {author} {\bibinfo {author} {\bibfnamefont {Izumi}\ \bibnamefont
  {Tomeno}}, \bibinfo {author} {\bibfnamefont {Hiromi~N.}\ \bibnamefont
  {Fuke}}, \bibinfo {author} {\bibfnamefont {Hitoshi}\ \bibnamefont {Iwasaki}},
  \bibinfo {author} {\bibfnamefont {Masashi}\ \bibnamefont {Sahashi}}, \ and\
  \bibinfo {author} {\bibfnamefont {Yorihiko}\ \bibnamefont {Tsunoda}},\
  }\bibfield  {title} {\enquote {\bibinfo {title} {{Magnetic neutron scattering
  study of ordered Mn3Ir}},}\ }\href {\doibase 10.1063/1.371298} {\bibfield
  {journal} {\bibinfo  {journal} {Journal of Applied Physics}\ }\textbf
  {\bibinfo {volume} {86}},\ \bibinfo {pages} {3853} (\bibinfo {year}
  {1999})}\BibitemShut {NoStop}%
\bibitem [{\citenamefont {Szunyogh}\ \emph {et~al.}(2009)\citenamefont
  {Szunyogh}, \citenamefont {Lazarovits}, \citenamefont {Udvardi},
  \citenamefont {Jackson},\ and\ \citenamefont
  {Nowak}}]{Szunyogh2009GiantPrinciples}%
  \BibitemOpen
  \bibfield  {author} {\bibinfo {author} {\bibfnamefont {L.}~\bibnamefont
  {Szunyogh}}, \bibinfo {author} {\bibfnamefont {B.}~\bibnamefont
  {Lazarovits}}, \bibinfo {author} {\bibfnamefont {L.}~\bibnamefont {Udvardi}},
  \bibinfo {author} {\bibfnamefont {J.}~\bibnamefont {Jackson}}, \ and\
  \bibinfo {author} {\bibfnamefont {U.}~\bibnamefont {Nowak}},\ }\bibfield
  {title} {\enquote {\bibinfo {title} {{Giant magnetic anisotropy of the bulk
  antiferromagnets IrMn and IrMn3from first principles}},}\ }\href {\doibase
  10.1103/PhysRevB.79.020403} {\bibfield  {journal} {\bibinfo  {journal}
  {Physical Review B - Condensed Matter and Materials Physics}\ }\textbf
  {\bibinfo {volume} {79}},\ \bibinfo {pages} {1--4} (\bibinfo {year}
  {2009})}\BibitemShut {NoStop}%
\bibitem [{\citenamefont {Szunyogh}\ \emph {et~al.}(2011)\citenamefont
  {Szunyogh}, \citenamefont {Udvardi}, \citenamefont {Jackson}, \citenamefont
  {Nowak},\ and\ \citenamefont {Chantrell}}]{Szunyogh2011AtomisticInterface}%
  \BibitemOpen
  \bibfield  {author} {\bibinfo {author} {\bibfnamefont {L.}~\bibnamefont
  {Szunyogh}}, \bibinfo {author} {\bibfnamefont {L.}~\bibnamefont {Udvardi}},
  \bibinfo {author} {\bibfnamefont {J.}~\bibnamefont {Jackson}}, \bibinfo
  {author} {\bibfnamefont {U.}~\bibnamefont {Nowak}}, \ and\ \bibinfo {author}
  {\bibfnamefont {R.}~\bibnamefont {Chantrell}},\ }\bibfield  {title} {\enquote
  {\bibinfo {title} {{Atomistic spin model based on a spin-cluster expansion
  technique: Application to the IrMn 3 /Co interface}},}\ }\href {\doibase
  10.1103/PhysRevB.83.024401} {\bibfield  {journal} {\bibinfo  {journal}
  {Physical Review B}\ }\textbf {\bibinfo {volume} {83}},\ \bibinfo {pages}
  {024401} (\bibinfo {year} {2011})}\BibitemShut {NoStop}%
\bibitem [{\citenamefont {Tsunoda}\ \emph {et~al.}(2007)\citenamefont
  {Tsunoda}, \citenamefont {Yoshitaki}, \citenamefont {Ashizawa}, \citenamefont
  {Mitsumata}, \citenamefont {Nakamura}, \citenamefont {Osawa}, \citenamefont
  {Hirono}, \citenamefont {Kim},\ and\ \citenamefont
  {Takahashi}}]{Tsunoda2007UncompensatedBilayers}%
  \BibitemOpen
  \bibfield  {author} {\bibinfo {author} {\bibfnamefont {M.}~\bibnamefont
  {Tsunoda}}, \bibinfo {author} {\bibfnamefont {S.}~\bibnamefont {Yoshitaki}},
  \bibinfo {author} {\bibfnamefont {Y.}~\bibnamefont {Ashizawa}}, \bibinfo
  {author} {\bibfnamefont {C.}~\bibnamefont {Mitsumata}}, \bibinfo {author}
  {\bibfnamefont {T.}~\bibnamefont {Nakamura}}, \bibinfo {author}
  {\bibfnamefont {H.}~\bibnamefont {Osawa}}, \bibinfo {author} {\bibfnamefont
  {T.}~\bibnamefont {Hirono}}, \bibinfo {author} {\bibfnamefont {D.~Y.}\
  \bibnamefont {Kim}}, \ and\ \bibinfo {author} {\bibfnamefont
  {M.}~\bibnamefont {Takahashi}},\ }\bibfield  {title} {\enquote {\bibinfo
  {title} {{Uncompensated antiferromagnetic spins at the interface in Mn–Ir
  based exchange biased bilayers}},}\ }\href {\doibase 10.1063/1.2710216}
  {\bibfield  {journal} {\bibinfo  {journal} {Journal of Applied Physics}\
  }\textbf {\bibinfo {volume} {101}},\ \bibinfo {pages} {09E510} (\bibinfo
  {year} {2007})}\BibitemShut {NoStop}%
\bibitem [{\citenamefont {Vallejo-Fernandez}\ \emph {et~al.}(2007)\citenamefont
  {Vallejo-Fernandez}, \citenamefont {Fernandez-Outon},\ and\ \citenamefont
  {O’Grady}}]{Vallejo-Fernandez2007MeasurementSystems}%
  \BibitemOpen
  \bibfield  {author} {\bibinfo {author} {\bibfnamefont {G.}~\bibnamefont
  {Vallejo-Fernandez}}, \bibinfo {author} {\bibfnamefont {L.~E.}\ \bibnamefont
  {Fernandez-Outon}}, \ and\ \bibinfo {author} {\bibfnamefont {K.}~\bibnamefont
  {O’Grady}},\ }\bibfield  {title} {\enquote {\bibinfo {title} {{Measurement
  of the anisotropy constant of antiferromagnets in metallic polycrystalline
  exchange biased systems}},}\ }\href {\doibase 10.1063/1.2817230} {\bibfield
  {journal} {\bibinfo  {journal} {Applied Physics Letters}\ }\textbf {\bibinfo
  {volume} {91}},\ \bibinfo {pages} {212503} (\bibinfo {year}
  {2007})}\BibitemShut {NoStop}%
\bibitem [{\citenamefont {Carpenter}\ \emph {et~al.}(2014)\citenamefont
  {Carpenter}, \citenamefont {Vick}, \citenamefont {Hirohata}, \citenamefont
  {Vallejo-Fernandez},\ and\ \citenamefont
  {O'grady}}]{Carpenter2014EffectNanostructures}%
  \BibitemOpen
  \bibfield  {author} {\bibinfo {author} {\bibfnamefont {R}~\bibnamefont
  {Carpenter}}, \bibinfo {author} {\bibfnamefont {A~J}\ \bibnamefont {Vick}},
  \bibinfo {author} {\bibfnamefont {A}~\bibnamefont {Hirohata}}, \bibinfo
  {author} {\bibfnamefont {G}~\bibnamefont {Vallejo-Fernandez}}, \ and\
  \bibinfo {author} {\bibfnamefont {K}~\bibnamefont {O'grady}},\ }\bibfield
  {title} {\enquote {\bibinfo {title} {{Effect of grain cutting in exchange
  biased nanostructures}},}\ }\href {\doibase 10.1063/1.4868328} {\bibfield
  {journal} {\bibinfo  {journal} {Citation: Journal of Applied Physics}\
  }\textbf {\bibinfo {volume} {115}},\ \bibinfo {pages} {17--905} (\bibinfo
  {year} {2014})}\BibitemShut {NoStop}%
\bibitem [{\citenamefont {Vallejo-Fernandez}\ \emph {et~al.}(2010)\citenamefont
  {Vallejo-Fernandez}, \citenamefont {Aley}, \citenamefont {Chapman},\ and\
  \citenamefont
  {O’Grady}}]{Vallejo-Fernandez2010MeasurementAntiferromagnets}%
  \BibitemOpen
  \bibfield  {author} {\bibinfo {author} {\bibfnamefont {G.}~\bibnamefont
  {Vallejo-Fernandez}}, \bibinfo {author} {\bibfnamefont {N.~P.}\ \bibnamefont
  {Aley}}, \bibinfo {author} {\bibfnamefont {J.~N.}\ \bibnamefont {Chapman}}, \
  and\ \bibinfo {author} {\bibfnamefont {K.}~\bibnamefont {O’Grady}},\
  }\bibfield  {title} {\enquote {\bibinfo {title} {{Measurement of the attempt
  frequency in antiferromagnets}},}\ }\href {\doibase 10.1063/1.3522887}
  {\bibfield  {journal} {\bibinfo  {journal} {Applied Physics Letters}\
  }\textbf {\bibinfo {volume} {97}},\ \bibinfo {pages} {222505} (\bibinfo
  {year} {2010})}\BibitemShut {NoStop}%
\bibitem [{\citenamefont {Craig}\ \emph {et~al.}(2008)\citenamefont {Craig},
  \citenamefont {Lamberton}, \citenamefont {Johnston}, \citenamefont {Nowak},
  \citenamefont {Chantrell},\ and\ \citenamefont
  {O’Grady}}]{Craig2008ABilayers}%
  \BibitemOpen
  \bibfield  {author} {\bibinfo {author} {\bibfnamefont {B.}~\bibnamefont
  {Craig}}, \bibinfo {author} {\bibfnamefont {R.}~\bibnamefont {Lamberton}},
  \bibinfo {author} {\bibfnamefont {A.}~\bibnamefont {Johnston}}, \bibinfo
  {author} {\bibfnamefont {U.}~\bibnamefont {Nowak}}, \bibinfo {author}
  {\bibfnamefont {R.~W.}\ \bibnamefont {Chantrell}}, \ and\ \bibinfo {author}
  {\bibfnamefont {K.}~\bibnamefont {O’Grady}},\ }\bibfield  {title} {\enquote
  {\bibinfo {title} {{A model of the temperature dependence of exchange bias in
  coupled ferromagnetic∕antiferromagnetic bilayers}},}\ }\href {\doibase
  10.1063/1.2830638} {\bibfield  {journal} {\bibinfo  {journal} {Journal of
  Applied Physics}\ }\textbf {\bibinfo {volume} {103}},\ \bibinfo {pages}
  {07C102} (\bibinfo {year} {2008})}\BibitemShut {NoStop}%
\bibitem [{\citenamefont {Callen}\ and\ \citenamefont
  {Callen}(1966)}]{Callen1966TheLaw}%
  \BibitemOpen
  \bibfield  {author} {\bibinfo {author} {\bibfnamefont {H.B.}\ \bibnamefont
  {Callen}}\ and\ \bibinfo {author} {\bibfnamefont {E.}~\bibnamefont
  {Callen}},\ }\bibfield  {title} {\enquote {\bibinfo {title} {{The present
  status of the temperature dependence of magnetocrystalline anisotropy, and
  the l(l+1)2 power law}},}\ }\href {\doibase 10.1016/0022-3697(66)90012-6}
  {\bibfield  {journal} {\bibinfo  {journal} {Journal of Physics and Chemistry
  of Solids}\ }\textbf {\bibinfo {volume} {27}},\ \bibinfo {pages} {1271--1285}
  (\bibinfo {year} {1966})}\BibitemShut {NoStop}%
\bibitem [{\citenamefont {Jenkins}\ and\ \citenamefont
  {Evans}(2018)}]{Jenkins2018EnhancedFilms}%
  \BibitemOpen
  \bibfield  {author} {\bibinfo {author} {\bibfnamefont {Sarah}\ \bibnamefont
  {Jenkins}}\ and\ \bibinfo {author} {\bibfnamefont {Richard F.~L.}\
  \bibnamefont {Evans}},\ }\bibfield  {title} {\enquote {\bibinfo {title}
  {{Enhanced finite size and interface mixing effects in iridium manganese
  ultra thin films}},}\ }\href {\doibase 10.1063/1.5038006} {\bibfield
  {journal} {\bibinfo  {journal} {Journal of Applied Physics}\ }\textbf
  {\bibinfo {volume} {124}},\ \bibinfo {pages} {152105} (\bibinfo {year}
  {2018})}\BibitemShut {NoStop}%
\bibitem [{\citenamefont {Neel}(1954)}]{Neel1954LapprocheMagnetostriction}%
  \BibitemOpen
  \bibfield  {author} {\bibinfo {author} {\bibfnamefont {Louis}\ \bibnamefont
  {Neel}},\ }\bibfield  {title} {\enquote {\bibinfo {title} {{L'approche
  {\`{a}} la saturation de la magn{\'{e}}tostriction}},}\ }\href {\doibase
  10.1051/jphysrad:01954001505037601} {\bibfield  {journal} {\bibinfo
  {journal} {J. Phys. Radium}\ }\textbf {\bibinfo {volume} {15}} (\bibinfo
  {year} {1954}),\ 10.1051/jphysrad:01954001505037601}\BibitemShut {NoStop}%
\bibitem [{\citenamefont {Alzate-Cardona}\ \emph {et~al.}(2019)\citenamefont
  {Alzate-Cardona}, \citenamefont {Sabogal-Su{\'{a}}rez}, \citenamefont
  {Evans},\ and\ \citenamefont
  {Restrepo-Parra}}]{Alzate-Cardona2019OptimalSystems}%
  \BibitemOpen
  \bibfield  {author} {\bibinfo {author} {\bibfnamefont {J~D}\ \bibnamefont
  {Alzate-Cardona}}, \bibinfo {author} {\bibfnamefont {D}~\bibnamefont
  {Sabogal-Su{\'{a}}rez}}, \bibinfo {author} {\bibfnamefont {R~F~L}\
  \bibnamefont {Evans}}, \ and\ \bibinfo {author} {\bibfnamefont
  {E}~\bibnamefont {Restrepo-Parra}},\ }\bibfield  {title} {\enquote {\bibinfo
  {title} {{Optimal phase space sampling for Monte Carlo simulations of
  Heisenberg spin systems}},}\ }\href {\doibase 10.1088/1361-648X/aaf852}
  {\bibfield  {journal} {\bibinfo  {journal} {Journal of Physics: Condensed
  Matter}\ }\textbf {\bibinfo {volume} {31}},\ \bibinfo {pages} {095802}
  (\bibinfo {year} {2019})}\BibitemShut {NoStop}%
\bibitem [{\citenamefont {Evans}\ \emph {et~al.}(2014)\citenamefont {Evans},
  \citenamefont {Fan}, \citenamefont {Chureemart}, \citenamefont {Ostler},
  \citenamefont {Ellis},\ and\ \citenamefont {Chantrell}}]{Evans2014}%
  \BibitemOpen
  \bibfield  {author} {\bibinfo {author} {\bibfnamefont {R.~F.L.}\ \bibnamefont
  {Evans}}, \bibinfo {author} {\bibfnamefont {W.~J.}\ \bibnamefont {Fan}},
  \bibinfo {author} {\bibfnamefont {P.}~\bibnamefont {Chureemart}}, \bibinfo
  {author} {\bibfnamefont {T.~A.}\ \bibnamefont {Ostler}}, \bibinfo {author}
  {\bibfnamefont {M.~O.A.}\ \bibnamefont {Ellis}}, \ and\ \bibinfo {author}
  {\bibfnamefont {R.~W.}\ \bibnamefont {Chantrell}},\ }\bibfield  {title}
  {\enquote {\bibinfo {title} {{Atomistic spin model simulations of magnetic
  nanomaterials}},}\ }\href {\doibase 10.1088/0953-8984/26/10/103202}
  {\bibfield  {journal} {\bibinfo  {journal} {Journal of Physics Condensed
  Matter}\ }\textbf {\bibinfo {volume} {26}} (\bibinfo {year} {2014}),\
  10.1088/0953-8984/26/10/103202}\BibitemShut {NoStop}%
\bibitem [{vmp(2019)}]{vmpr}%
  \BibitemOpen
  \bibfield  {title} {\enquote {\bibinfo {title} {{\textsc{vampire} software
  package v5 available from https://vampire.york.ac.uk/}},}\ }\href@noop {} {\
  (\bibinfo {year} {2019})}\BibitemShut {NoStop}%
\bibitem [{\citenamefont {Asselin}\ \emph {et~al.}(2010)\citenamefont
  {Asselin}, \citenamefont {Evans}, \citenamefont {Barker}, \citenamefont
  {Chantrell}, \citenamefont {Yanes}, \citenamefont {Chubykalo-Fesenko},
  \citenamefont {Hinzke},\ and\ \citenamefont
  {Nowak}}]{Asselin2010ConstrainedAnisotropy}%
  \BibitemOpen
  \bibfield  {author} {\bibinfo {author} {\bibfnamefont {P.}~\bibnamefont
  {Asselin}}, \bibinfo {author} {\bibfnamefont {R.~F.~L.}\ \bibnamefont
  {Evans}}, \bibinfo {author} {\bibfnamefont {J.}~\bibnamefont {Barker}},
  \bibinfo {author} {\bibfnamefont {R.~W.}\ \bibnamefont {Chantrell}}, \bibinfo
  {author} {\bibfnamefont {R.}~\bibnamefont {Yanes}}, \bibinfo {author}
  {\bibfnamefont {O.}~\bibnamefont {Chubykalo-Fesenko}}, \bibinfo {author}
  {\bibfnamefont {D.}~\bibnamefont {Hinzke}}, \ and\ \bibinfo {author}
  {\bibfnamefont {U.}~\bibnamefont {Nowak}},\ }\bibfield  {title} {\enquote
  {\bibinfo {title} {{Constrained Monte Carlo method and calculation of the
  temperature dependence of magnetic anisotropy}},}\ }\href {\doibase
  10.1103/PhysRevB.82.054415} {\bibfield  {journal} {\bibinfo  {journal}
  {Physical Review B}\ }\textbf {\bibinfo {volume} {82}},\ \bibinfo {pages}
  {054415} (\bibinfo {year} {2010})}\BibitemShut {NoStop}%
\bibitem [{\citenamefont {Ellis}\ \emph {et~al.}(2015)\citenamefont {Ellis},
  \citenamefont {Evans}, \citenamefont {Ostler}, \citenamefont {Barker},
  \citenamefont {Atxitia}, \citenamefont {Chubykalo-Fesenko},\ and\
  \citenamefont {Chantrell}}]{Ellis}%
  \BibitemOpen
  \bibfield  {author} {\bibinfo {author} {\bibfnamefont {M.~O.~A.}\
  \bibnamefont {Ellis}}, \bibinfo {author} {\bibfnamefont {R.~F.~L.}\
  \bibnamefont {Evans}}, \bibinfo {author} {\bibfnamefont {T.~A.}\ \bibnamefont
  {Ostler}}, \bibinfo {author} {\bibfnamefont {J.}~\bibnamefont {Barker}},
  \bibinfo {author} {\bibfnamefont {U.}~\bibnamefont {Atxitia}}, \bibinfo
  {author} {\bibfnamefont {O.}~\bibnamefont {Chubykalo-Fesenko}}, \ and\
  \bibinfo {author} {\bibfnamefont {R.~W.}\ \bibnamefont {Chantrell}},\
  }\bibfield  {title} {\enquote {\bibinfo {title} {The landau–lifshitz
  equation in atomistic models},}\ }\href {\doibase 10.1063/1.4930971}
  {\bibfield  {journal} {\bibinfo  {journal} {Low Temperature Physics}\
  }\textbf {\bibinfo {volume} {41}},\ \bibinfo {pages} {705--712} (\bibinfo
  {year} {2015})},\ \Eprint
  {http://arxiv.org/abs/https://doi.org/10.1063/1.4930971}
  {https://doi.org/10.1063/1.4930971} \BibitemShut {NoStop}%
\end{thebibliography}%

\end{document}